\documentclass[letterpaper, 10 pt, conference]{ieeeconf}  
\usepackage{amsmath}
\usepackage{tikz}
\usepackage{subcaption}
\usetikzlibrary{positioning}
\usetikzlibrary{arrows}
\usetikzlibrary{backgrounds}
\usetikzlibrary{calc}
\usetikzlibrary{shapes}
\usetikzlibrary{automata}
\usetikzlibrary{shadows}
\usetikzlibrary{decorations.pathreplacing}

\IEEEoverridecommandlockouts                              

\title{\LARGE \bf
Synthesis and Implementation of Distributed Supervisory Controllers with Communication Delays
}

\author{R.H.J. Schouten, L. Moormann$^{1}$, J.M. van de Mortel-Fronczak$^{1}$, and J.E. Rooda$^{1}$%
\thanks{*This work was supported by Rijkswaterstaat, part of the Dutch ministry of infrastructure and water management}%
\thanks{$^{1}$Department of Mechanical Engineering,
        University of Technology Eindhoven, The Netherlands
        {\tt\small l.moormann@tue.nl}}%
}

\RequirePackage{relsize}

\usetikzlibrary{shapes.arrows}
\usetikzlibrary{shapes.misc}

\definecolor{evalcolor}{rgb}{0.000,0.400,0.800}

\providecolor{bg}{named}{white}

\tikzset{
fmbe/.style= {
  , x = 6.5em
  , y = 6.5em
  , node distance = 7em
  , on grid
  , shorten >=1pt
  , every artifact/.style = { draw
                            , minimum size=2.5em
                            , font=\larger
                            }
  , document/.style={ every artifact
                    , shape=document 
                    }
  , model/.style={ every artifact
                 , circle
                 , inner sep=0pt
                 }
  , realization/.style={ every artifact
                       , rectangle
                       }
  , infra/.style = { draw
                   , double arrow
                   , double arrow head extend=1ex
                   }
  , actlb/.style = { above
                   , inner sep=0pt
                   , sharp corners
                   , font=\strut
                   }
  , actlb'/.style = { actlb
                    , anchor=mid
                    , fill=bg
                    , font=
                    , inner sep=2pt
                    }
  , act/.style = { ->
                 , rounded corners
                 , every node/.style = actlb
                 }
  , eval/.style = { 
                  , ->
                  , rounded corners
                  , every node/.style = actlb
                  , dashed
                  , draw
                  , evalcolor
                  , font=\strut
                  , inner sep=0pt
                  }
  , hidden/.style = { 
                    , opacity=.5
                    , every node/.append style={fill opacity=1,text opacity=.5}
                    }
  }
}

\pgfdeclareshape{document}{
    \inheritsavedanchors[from=rectangle] 
    \inheritanchorborder[from=rectangle]
    \inheritanchor[from=rectangle]{center}
    \inheritanchor[from=rectangle]{mid}
    \inheritanchor[from=rectangle]{base}
    \inheritanchor[from=rectangle]{north}
    \inheritanchor[from=rectangle]{south}
    \inheritanchor[from=rectangle]{west}
    \inheritanchor[from=rectangle]{east}
    \inheritanchor[from=rectangle]{mid west}
    \inheritanchor[from=rectangle]{base west}
    \inheritanchor[from=rectangle]{north west}
    \inheritanchor[from=rectangle]{south west}
    \inheritanchor[from=rectangle]{mid east}
    \inheritanchor[from=rectangle]{base east}
    \inheritanchor[from=rectangle]{north east} 
    \inheritanchor[from=rectangle]{south east}
    \backgroundpath{
        \southwest \pgf@xa=\pgf@x \pgf@ya=\pgf@y
        \northeast \pgf@xb=\pgf@x \pgf@yb=\pgf@y
        \pgf@xc=\pgf@xb \advance\pgf@xc by-5pt 
        \pgf@yc=\pgf@yb \advance\pgf@yc by-5pt
        \pgfpathmoveto{\pgfpoint{\pgf@xa}{\pgf@ya}}
        \pgfpathlineto{\pgfpoint{\pgf@xa}{\pgf@yb}}
        \pgfpathlineto{\pgfpoint{\pgf@xc}{\pgf@yb}}
        \pgfpathlineto{\pgfpoint{\pgf@xb}{\pgf@yc}}
        \pgfpathlineto{\pgfpoint{\pgf@xb}{\pgf@ya}}
        \pgfpathclose
        \pgfpathmoveto{\pgfpoint{\pgf@xc}{\pgf@yb}}
        \pgfpathlineto{\pgfpoint{\pgf@xc}{\pgf@yc}}
        \pgfpathlineto{\pgfpoint{\pgf@xb}{\pgf@yc}}
        \pgfpathlineto{\pgfpoint{\pgf@xc}{\pgf@yc}}
    }
}

\begin{document}

\maketitle
\thispagestyle{empty}
\pagestyle{empty}

\begin{abstract}
In literature, extensive research has been done with respect to synthesis of supervisory controllers. Such synthesized supervisors can be distributed for implementation on multiple physical controllers. This paper discusses a method for distributing a synthesized supervisory controller. In this method, dependency structure matrices are used to distribute a system, the supervisor is then distributed accordingly, using existing localization theory. The existence of communication delays between supervisory controllers is unavoidable in a distributed application. The influence of these delays on the behavior of a supervisor is therefore studied using delay robustness theory. This paper introduces the use of mutex algorithms to make the distributed supervisor delay-robust. A case study is used to demonstrate the method and hardware in the loop testing is used to validate the resulting distributed supervisor. 

\end{abstract}

\section{INTRODUCTION}
Discrete Event Systems (DES) \cite{Cassandras2008} can be found in a range of domains. Examples are manufacturing systems, infrastructural systems and logistic systems. The behavior of such systems is described by discrete events, for example a button is pushed or lamp is switched on. In order to assure only the desired behavior of the system occurs, a controller is needed. Such a controller is called a supervisory controller. Ramadge and Wonham introduced Supervisory Control Theory (SCT) in \cite{Ramadge1982}, which provides a method to synthesize such a supervisory controller. This method uses a model of the uncontrolled system, which describes what the system can do, and a model of the requirements, which states what the system is allowed to do.

Synthesized supervisory controllers can be implemented on a physical platform, for example a Programmable Logic Controller (PLC). The memory of such a PLC is limited, therefore it is sometimes not possible to control a large scale system with a single controller.  Using multiple PLCs can raise the total memory capacity, however a distributed supervisory controller is required for such an implementation. A system is then not controlled by a single global controller, but by multiple local controllers that are able to communicate with one another. Using a distributed supervisory controller implemented on multiple PLCs instead of a supervisory controller implemented on a single PLC might reduce PLC cycle times, as the computing power can be increased and the size of the controller can be reduced. It should be noted, however, that in a distributed implementation, communication might be required, which can increase the total required memory and computation power.

This paper discusses a method to synthesize and implement a distributed supervisory controller while taking into account communication delays. The paper is structured as follows. First the existing relevant theory from literature is discussed concerning the topics of SCT, distributed supervisory control, Dependency Structure Matrices (DSM), localization, and mutual exclusion algorithms which are used to counteract the effects of communication delays on the order in which events occur. Next, a case study is introduced in Section \ref{sec:usecase}. The method introduced in this paper is then discussed in two main steps. First, DSMs are used to distribute the uncontrolled system, i.e. the plant. Second, the acquired distribution of the system is used to distribute the supervisor for that system accordingly. In Section \ref{sec:HIL}, hardware in the loop testing is used to validate the distributed supervisor for the case study. Final remarks and conclusions are given in Section \ref{sec:Conclusions}. 

\section{PRELIMINARIES}
\subsection{Supervisory Control Theory}


In the context of SCT, a supervisor can be synthesized from a model of the plant, i.e. the uncontrolled system, and a model of the requirements. SCT notations and definitions as discussed in this section are taken from \cite{Cassandras2008} and \cite{Wonham2015}. In this paper, DES are modeled using Finite Automata (FA). A FA is defined by a 5-Tuple: $\left(X, \Sigma, \xi, x_{0}, X_{m} \right)$. Here, $X$ denotes a finite set of states and $\Sigma$ denotes a finite set of events. Event set $\Sigma$ is partitioned into two disjoint sets $\Sigma_c$ and $\Sigma_u$, controllable events and uncontrollable events, respectively. Controllable events may be disabled by a supervisor, whereas uncontrollable events may not. Furthermore, $\xi : X \times \Sigma \rightarrow X $ is the transition relation for events $\sigma \in \Sigma$. $x_{o}$ denotes the initial state and $X_{m}$ the set of marked states. 

The set of all finite strings of events over $\Sigma$, including the empty string $ \epsilon$, is denoted by  $\Sigma^*$. The set of possible strings in FA $G$ is referred to as its language $L\left(G\right)$. Formally: $L \left( G \right) = \{s|s \in \Sigma^*, \xi(x_0, s)!\}$, where $\xi(x_0, s)!$ denotes that the string $s$ of events leads to a state in $X$ from the initial state $x_0$. The set of strings leading to a marked stated is known as the marked language of $G$ denoted as $L_m(G)$, or more formally: $L_m \left( G \right) = \{s|s \in \Sigma^*, \xi(x_0, s)\in X_m\}$. Note that based on the given definitions $	L_m \left( G \right) \subseteq	L \left( G \right)$. 


A projection is a frequently used opreation in SCT. For the definition of projections, proposed by \cite{Cassandras2008}, consider event set $\Sigma_s \in \Sigma$ of FA $G$. The projection $P_s$, from $\Sigma^*$ to $\Sigma_s^*$, replaces all events in $\Sigma \setminus \Sigma_s$ by $\epsilon$. Another frequently used operation is the synchronous product, denoted by $||$ . The synchronous product of FAs $G_1$ and $G_2$, is	$G_1 || G_2 = \\
	(X_1 \times X_2, \Sigma_1 \bigcup \Sigma_2, \xi, (x_{0,1}, x_{0,2}), X_{m,1} \times X_{m,2})$, where:

 $\vspace{-25pt}$
 
\begin{equation*}
    \resizebox{\hsize}{!}{%
        $
        \xi\left(\left(x_{1}, x_{2}\right), e\right):=\\
		\\
		\left\{\begin{array}{lll}
			\left(\xi_{1}\left(x_{1}, e\right), \xi_{2}\left(x_{2}, e\right)\right) & \text { if } \xi_1(x_1,e)! & \wedge \ \xi_2(x_2,e)! \\
			\left(\xi_{1}\left(x_{1}, e\right), x_{2}\right) & \text { if } \xi_1(x_1,e)!  & \wedge \ e\notin\Sigma_2 \\ 
			\left(x_{1}, \xi_{2}\left(x_{2}, e\right)\right) & \text { if } e \notin \Sigma_1 & \wedge \ \xi_2(x_2,e)! \\
			\text { undefined } & \text { otherwise } &
		\end{array}\right.
        $%
	}
\end{equation*}

The DES discussed in this paper are composed systems. A composed system $G$ is defined by the synchronous product of a set of subplants $G_i$, with $i \in I = (1,2,...,n)$, i.e. $G = ||_{i \in I}G_i$. A product system is defined by \cite{Ramadge1989} as a composed system where all subplants have disjoint alphabets. In \cite{Queiroz2000}, the Most Refined Product System (MRPS) of a composed system is defined as the product system with the largest number of subplants that do not share events. A composed system can be transformed into a MRPS using the procedure in \cite{DeQueiroz2000}.

The supervisor synthesis procedure uses a plant model and a requirement model, to derive a supervisor. This supervisor disables controllable events, to ensure the system under supervision satisfies a number of properties. These properties are stated by \cite{Ramadge1989} as safety, nonblockingness, controllability and maximal permissiveness. 

Safety describes that the supervisor meets all the requirements stated by the requirement model. Safety does not state the correctness of the requirements, it only states that the supervised system meets the requirements. A supervised system that is nonblocking can always reach a marked state from any state reachable from the initial state, or formally, any reachable state is coreachable. Nonblockingness is confirmed if the prefix closure $\overline{L_{m}(S)}=L(S)$ \cite{Cai2009}. Controllability means that no uncontrollable events are disabled by the supervisor. A maximally permissive supervisor allows all behavior that satisfies the properties safety, nonblockingness and controllability. 

The supervisor synthesis procedure described in \cite{Ramadge1982} results in a monolithic supervisor where all requirements are captured. When the complexity of the system increases so does the computational complexity of synthesizing a monolithic supervisor. Different synthesis methods (or supervisor architectures) have been proposed in literature to decrease the computation times of synthesizing supervisors. 

The first alternative to monolithic synthesis is modular synthesis as proposed in \cite{Wonham1988}. In modular synthesis, a local supervisor for each requirement in the requirement model is derived. For large scale systems, synthesis of local supervisors is computationally less complex than for a single monolithical supervisor. The conditions for the existence of a nonblocking modular supervisor are stronger than for a monolithic solution, as stated by \cite{Wonham1988}. Namely to assure nonblockingness of the global supervisor, local supervisors need to be non-conflicting. \cite{Wonham1988} defines conditions for local supervisors such that they are non-conflicting. If the local supervisors are non-conflicting, the global supervisor is the synchronous product of all the local supervisors.

Decentralized supervisors, as discussed in \cite{Yoo2002}, are similar to modular supervisors. Each decentralized supervisor observes a subset and controls a subset of the event set of the system. For a system with two supervisors, the global observable set is defined as $\Sigma_{o}=\Sigma_{o,1} \cup \Sigma_{o,2}$, where $\Sigma_{o,i}$ the set of events observable by supervisor $i$ and the global controllable set is defined as $\Sigma_{c}=\Sigma_{c,1} \cup \Sigma_{c,2}$, where $\Sigma_{c,i}$ the set of events controllable by supervisor $i$. The supervisors are fused into one supervisor using a fusion rule. The supervisors can be fused by intersection, union or a combination of both. If local supervisors are defined such that events which are not in their controllable event set, i.e. $\Sigma_{c} \backslash \Sigma_{c, i}$, are always enabled, intersection is used. If these events are always disabled, union is used. If $\Sigma_{c}$ is partitioned into a set events $\Sigma_{c,e}$ that are enabled by default and a set $\Sigma_{c,d}$ that are disabled by default, both union and intersection are used. It should be noted that a decentralized supervisor is not the same as a distributed supervisor. A decentralized supervisor is implemented as a single supervisory controller, whereas a distributed supervisor is implemented as multiple supervisory controllers that are able to communicate. 

More recently a multilevel architecture has been introduced in \cite{Komenda2016}. The multilevel architecture uses a MultiLevel Discrete Event System (MLDES) which consists of a tree-based structure. For each node in the tree, a local supervisor that influences a set of subplants and satisfies a subset of requirements is synthesized. The set of local supervisors is nonblocking, safe, and controllable, however, maximal permissiveness is not guaranteed in multilevel synthesis. 

\subsection{Distributed supervisors}
In literature, several contributions to the development of the distribution of supervisory controllers over multiple physical controllers have been made. \cite{Hanisch2005} models a distributed controller using petri nets. Petri nets are a modelling formalism alternative to FA. This controller is verified using the SESA model checker. No synthesis is done, however, the paper provides useful insights in distributing controllers and the required communication between controllers. \cite{Zgorzelski2016} discusses synchronization of events in DES using networked controllers. Networked controllers are similar to distributed controllers in the sense that multiple local controllers communicate to apply control action to a plant. However, in networked control each controller controls a separate plant. The separate plants may have physical couplings. The developed method is used to enable multiple controllers to perform a synchronized task, for example two robot arms picking up an object together. The controllers in this case are autonomous and communicate with one another to perform the task synchronously.

There have also been contributions in literature towards acquiring distributed controllers through synthesis. \cite{Reijnen2020} discusses a method for synthesizing and implementing a supervisory controller for safety PLCs. In safety PLCs, the controller consists of a regular and a safety controller, which communicate through internal data buffers. The controllers are updated alternately, i.e. in a synchronized scan cycle. This method is not used in this paper, as it is expected that this solution is not scalable for larger numbers of controllers, because the synchronized way of updating controllers will introduce increasingly large update delays. Moreover, this method of splitting a supervisor is designed specifically for a system with a set of regular and safety requirements. It is unclear how to translate this to a more general distributed synthesis method.  

\cite{Su2010} introduces an aggregative method for synthesizing a distributed supervisor. There is no need for synthesizing a global supervisor (i.e. a supervisor for the entire plant and all its requirements) in this method, as local supervisors are synthesized right away. Computation times can therefore be relatively low. This is a great advantage in the scalability of the method. In order to achieve global nonblockingness, abstractions of other local supervisors are used during the synthesis of a local supervisor. Therefore the order in which local supervisors are synthesized is important for both the computational complexity of the supervisor as well as for the existence of a supervisor. \cite{Su2010} gives some guidelines to choose an efficient ordering, however, it is unclear how to find an optimal ordering. Moreover, the computations that are needed are relatively complex.


In contrast to the bottom-up approach of aggregate synthesis, supervisor localization as done by \cite{Cai2009} is a top-down approach to acquiring distributed supervisors. First a global supervisor for the entire plant is synthesized. A localized version of this global supervisor is then implemented for each component or component group. The provided method is a relatively simple way of obtaining a distributed supervisor. The behavior of this distributed supervisor is equal to that of the global supervisor. The concept of localization is further discussed in Section \ref{sec:locTheory}. 

Both the aggregative method of \cite{Su2010} and the localization method of \cite{Cai2009} do not discuss the effects of communication delays between local supervisors. \cite{Rashidinejad2018} introduces a synthesis method for (non-distributed) supervisors with a known delay on its inputs and outputs. The effects of communication delays can be similar for distributed supervisors, therefore \cite{Rashidinejad2018} gives some useful insights in the effects of communication delays. In \cite{Kalyon2011}, a distributed supervisor synthesis method is proposed which takes into account the effects of communication delays. \cite{Kalyon2011} states that more research is needed to reduce the amount of communication required. This method is therefore not used in this paper. \cite{Zhang2016} gives further insight in the effects of communication delays for distributed supervisors created by localization. This topic is further discussed in Section \ref{sec:delayrobust}. 


\subsection{Dependency Structure Matrix}
\label{sec:DSM}
Dependency Stucture Matrices (DSM), as introduced by \cite{Eppinger2012}, provide a means to give insight in the dependencies within a process or system. A DSM is a square matrix where the components of a system are the elements of both the horizontal and vertical axis. Each entry in the matrix denotes a dependency between two components. \cite{Goorden2018} shows that DSMs can be used for DES, to give insight in the dependencies between components within DES. A plant model and a requirement model are needed for the creation of such a DSM. The components of the plant model are listed on the axes of the DSM. Through requirements in the requirement model, different components depend on each other. These dependencies are shown in the DSM. Other types of dependencies, for example spatial dependencies, can also be shown in a DSM. 

In a DSM, clusters of closely connected components can be identified. When clustering components in a DSM, the order in which the components are placed on the axes is changed and components within a cluster are grouped. The aim is to identify clusters in such a way that the number of dependencies between components within a cluster is maximized and dependencies between components of different clusters is minimized. In literature, extensive research has been done to provide efficient clustering algorithms as K-means \cite{MacQueen1967} or spectral clustering \cite{Capocci2005}, or by using more general optimization methods, for example genetic algorithms \cite{Yu2007} or simulated annealing \cite{Thebeau2001}.

%

Clustering DSMs can also be done using Markov clustering, which is shown by \cite{Wilschut2018}. Markov clustering can create multiple hierarchical levels and can detect bus elements. Bus elements are elements in the DSM that have a large number of dependencies. Markov clustering is a relatively complex algorithm that uses the four clustering parameters $\alpha$, $\beta$, $\mu$ and $\gamma$ to tune the clustering of a DSM. Parameter $\gamma$ is a threshold value for the number of detected bus elements, increasing $\gamma$ decreases the number of bus elements. $\beta$ and $\mu$ are both used to tune the cluster size and the number of hierarchical levels. $\alpha$ has less influence on the clustering results \cite{Dongen2008}. \cite{Wilschut2018} shows that Markov clustering is a scalable and versatile solution for clustering DSMs.
%

\subsection{Localization}
\label{sec:locTheory}
Supervisor localization is introduced by \cite{Cai2008} for FA. It takes a global supervisor $\textbf{SUP} = (X, \Sigma, \xi , x_0,X_m)$, which is defined for a global plant $G = (Y, \Sigma, \eta, y_0, Y_m)$. Plant $G$ consists of component agents $G_k$ defined over disjoint alphabets $\Sigma^k, \ k \in K$ with $K$ an index set, where $\Sigma=\bigcup\left\{\Sigma^{k} \mid k \in K\right\}$. The global supervisor $\textbf{SUP}$ is localized into local supervisors $\textbf{LOC}_k$ for component agents $G_k$.

$\textbf{LOC}_k$ is a copy of $\textbf{SUP}$, where the controllable event set is adjusted to $\Sigma^k_c = \Sigma_{c} \bigcap \Sigma^k$. Events $\Sigma \setminus \Sigma^k_c$ are uncontrollable in $\textbf{LOC}_k$. All events in $\Sigma_{c}$ are controllable in exactly one local supervisor. The observable event set of $\textbf{LOC}_k$ is not adjusted, i.e. a local supervisor can observe events that are controlled by another local supervisor. Therefore, multiple local supervisors might interact to observe these so called shared events. $\textbf{LOC}_k$ is defined such that it observes events in $\Sigma^k$ directly from the plant. Events $\Sigma \setminus \Sigma^k$ are not observed directly from the plant, but can be observed in other local supervisors through communication between local supervisors. In \cite{Cai2008}, it is assumed no communication delays occur.

Using a monolithic global supervisor as a starting point for localization is often suboptimal as localization has to be done for a relatively large supervisor. \cite{Cai2010} shows that for large scale systems it is possible to use a more efficient method. In \cite{Cai2010}, a global supervisor is synthesized in a decentralized manner as discussed in \cite{Feng2009}, i.e. the global supervisor is the synchronous product of a number of decentralized supervisors and coordinators for nonblockingness. Instead of localizing the global supervisor for each component agent, only the decentralized supervisors that control events for a component agent are localized for that agent. It is possible that multiple decentralized supervisors control events in a component agent. In this case each of these supervisors is localized. It is also possible that one decentralized supervisor disables events in multiple component agents, in that case the decentralized supervisor is localized for each of those agents. The resulting local supervisors impose global nonblockingness, safety, and controllability, if the original global decentralized supervisor also imposed the same properties. 

\begin{figure}[h]
	\centering
	\includegraphics[width = .8\linewidth]{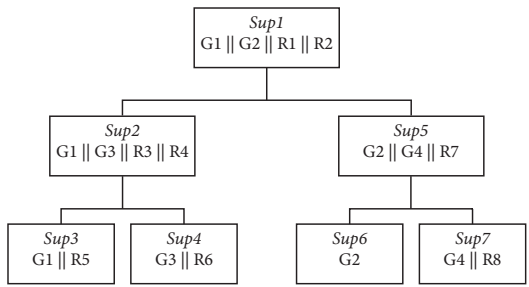}
	\caption{Multilevel supervisor example}
	\label{fig:MLDES_example}
\end{figure}

Similar to a decentralized supervisor, a multilevel supervisor can be used. A supervisor obtained by multilevel synthesis as described in \cite{Goorden2018}, consists of a number of supervisors that disable events in subsets of agents. For an example plant $G$ consisting of 4 component agents $G_k \ (k= 1,2,3,4)$, with a requirement model consisting of 8 requirements $R_i \ (i=1,2,...,8)$, multilevel synthesis is performed. The resulting MLDES tree is given in Fig. \ref{fig:MLDES_example}. The global multilevel supervisor is given by $S = ||_i \  Sup_i$, with $i = 1, 2,...,7$. Note tat $Sup6$ does not contain any requirements, any requirements that refer to component $G2$ are already in $Sup1$ and $Sup5$. 

\begin{figure*}[h]
	\centering
	\includegraphics[width = .8\linewidth]{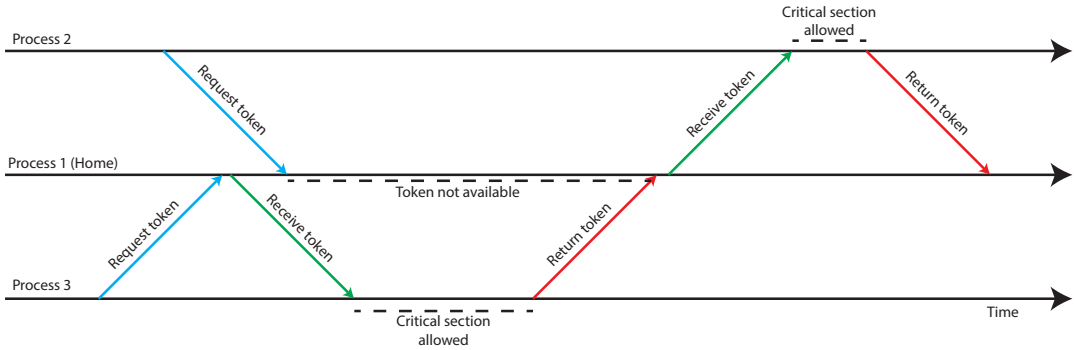}
	\caption{Home-based token passing mutual exclusion}
	\label{fig:Home_based}
\end{figure*}

The multilevel supervisor can now be localized for multiple component agents or groups of component agents. If for example a distribution of two agent groups is chosen, group 1 containing $G1$ and $G3$ and group 2 containing $G2$ and $G4$, it is not necessary to localize the entire multilevel supervisor. As group 1 is only controlled by supervisors $Sup1$, $Sup2$, $Sup3$ and $Sup4$ as depicted in Figure \ref{fig:MLDES_example}, only these supervisors need to be localized. Local supervisor $\textbf{LOC}_1$ is the set of localized versions of supervisors $Sup1$, $Sup2$, $Sup3$ and $Sup4$. Similarly, local supervisor $\textbf{LOC}_2$, for group 2, is the set of localized versions of supervisors $Sup1$, $Sup5$, $Sup6$ and $Sup7$. $\textbf{LOC}_1$ now controls agents $G1$ and $G3$ while observing $G2$, whereas $\textbf{LOC}_2$ now controls agents $G2$ and $G4$ while observing $G1$.

In \cite{Cai2009}, a localization algorithm is presented that creates an abstraction of the local supervisors, with the aim of reducing the complexity of the local supervisors such that only behavior relevant to the corresponding component agent is observed. Creating the abstraction improves the local supervisors in the sense that the statespace of the supervisor reduced. Moreover, the number of shared events and therefore the required amount of communication is reduced. In this paper, the algorithm is not used, as the local supervisors derived from a multilevel supervisor, are already relatively small. The unabstracted local supervisors are used instead. 

\subsection{Mutual exclusion algorithms}
\label{sec:mutextheory}
Mutual exclusion algorithms pose a solution to the problem in concurrent programming where multiple processes try to enter a, so-called, `critical section' of their program simultaneously, while this is not allowed. Many solutions have been proposed since the presentation of the Dijkstra's algorithm in \cite{Dijkstra1965}. \cite{Lamport1978} first introduced a solution to the distributed mutual exclusion problem, where the processes do not share any memory.

In most applications, mutual exclusion algorithms must ensure liveness, safety, and nonblockingness, as discussed in \cite{Kanrar2018}. Liveness, sometimes referred to as starvation freedom, implies that a process requesting to enter its critical section is eventually allowed to do so. Safety implies that no two processes can enter their critical section simultaneously. An algorithm is nonblocking if it is deadlock free. Deadlock is the situation where, in the algorithm, no action can be taken anymore, from the first moment of deadlock onward. Note that the definitions of nonblockingness for mutual exclusion algorithms and FA are similar, but not fully the same. An algorithm is considered fair, if the first process to request access is granted access the first. Some algorithms are able to prioritize one process over another. Most algorithms aim to reduce the amount of communication, i.e. the number of messages required to get access to a critical section.

Distributed mutual exclusion algorithms typically follow either a token-based (as for example in \cite{Raymond1989}) or a permission-based approach (as for example in \cite{Ricart1981}). Token-based algorithms, as defined in \cite{Raynal2013}, use a mobile object, referred to as the token, which travels from process to process. The process in possession of the token is allowed to enter its critical section. As there is only one token, only one process is allowed to enter its critical section simultaneously. Different network structures have been proposed, as discussed in \cite{Kanrar2018}, in efforts to reduce the amount of communication. In permission-based algorithms, a process requests permission to enter its critical section from all other processes or from a subgroup of other processes called a quorum or coterie. \cite{Kanrar2018} states that generally token-based algorithms need less communication than permission-based algorithms. Permission-based algorithms are generally more suited for implementation of fairness or prioritization. 

A simple implementation for a mutual exclusion algorithm is a home-based token algorithm as discussed in \cite{Raynal2013}. In this algorithm, there exists one home process and a number of other processes. The token normally resides at the home process. Any process can request the token from the home process and the home process will then send the token the requesting process. When the requesting process leaves the critical section it returns the token to the home process. If a process requests the token while the token is already in use by some other process, the requesting process will be added to a queue in the home process. This queue is a simple FIFO queue. The home process will always send the token to the first process in the queue as soon as possible. A process cannot enter the queue again if it is already in the queue. This algorithm requires little communication as each process only communicates with the home process. The process becomes inefficient if a large number of processes has to use the token often and for a short amount of time, as the token needs to be returned to the home process after every use instead of moving to the next process right away.

Figure \ref{fig:Home_based} shows an example case for a home-based token passing algorithm. Process 1 is the home process which initially holds the token. Process 3 requests the token from the home process and receives it. Process 3 is now allowed to enter the critical section. Next process 2 requests the token, the token however is not available as the token is held by process 3. Process 1 adds process 2 to the queue. As process 3 finishes using the critical section it returns the token to the home process. The home process now checks the queue and sends the token to process 2. Process 2 is now allowed to enter the critical section and returns the token afterwards.

\section{Case study: Pump-cellar system for tunnels}
\label{sec:usecase}
\begin{figure}[h]
	\centering
	\includegraphics[width=.75\linewidth]{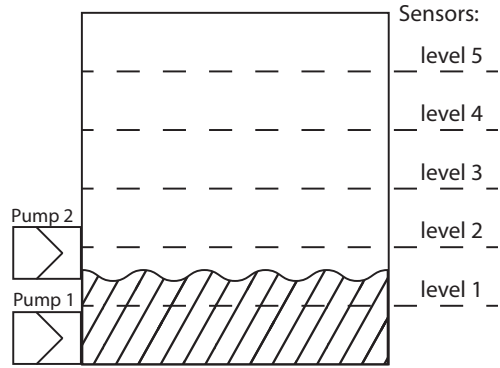}
	\caption{Pump cellar}
	\label{fig:pompkelder}	
\end{figure}

Throughout the remainder of this paper a case study is used to show an implementation of the discussed method. The case study concerns a pump-cellar system which is used in traffic tunnels to collect rain water. In this case, the system contains 3 cellars and each of these cellars is equipped with 2 pumps and 5 sensors, as depicted in Fig. \ref{fig:pompkelder}. The 5 sensors indicate when the water in the cellar has reached a certain level and the pumps are used to transfer water out of the pump-cellar. There are 2 main pump-cellars, which can pump water out of the system and 1 middle pumpceller, which can pump water to either of the 2 main pump-cellars. The CIF3 \cite{Beek2014} modeling language is used to model the system. 

\subsection{Plant model}

Each pump-cellar has a mode in which it operates, the mode can be $\mathit{Store}$, $\mathit{Empty}$ or $\mathit{Off}$. Based on the mode of the tunnel the pumps should be on or off at certain water levels. The mode of the tunnel can be changed manually or automatically. Manual control is done through the use of 3 buttons. Automatic changes can be made based on the state of the tunnel. The tunnel consists of 2 tubes, for which the automaton is shown in Fig. \ref{fig:SV} (event names are abbreviated here). Each of these tubes normally is in the state $\mathit{Operational}$. In case of an emergency, due to for example a car crash or fire, the state of each of these tubes can be changed to $\mathit{Emergency}$. After the emergency is solved the state is changed to $\mathit{Recovery}$ and finally back to $\mathit{Operational}$. Finally, the control mode, automatic or manual, is changed using 2 buttons. 

\begin{figure}[h]  
	\centering 
	\begin{minipage}{0.49\linewidth}
		\begin{center}
	\begin{tikzpicture}[auto,->,node distance=2cm,align=center,font=\smaller\it, uncontrollable/.style={densely dashed}]
	\node[initial left, initial text = {}, state,accepting,minimum size=.5cm, label=below:{Off}] (0) {};
	\node[state,right of=0,minimum size=.5cm, label=below:{On}] (1) {};
	
	\path[->,thick] (0) edge [bend left] node {\it c\_on} (1);
	\path[->,thick] (1) edge [bend left] node {\it c\_off} (0);
	\end{tikzpicture}
\end{center}
		\caption{Pump automaton} \label{fig:pump}  
	\end{minipage}
	\begin{minipage}{0.49\linewidth}
		\begin{center}
	\begin{tikzpicture}[auto,->,node distance=2cm,align=center,font=\smaller\it, uncontrollable/.style={densely dashed}]
	\node[initial left, initial text = {}, state,accepting,minimum size=.5cm, label=below:{Off}] (0) {};
	\node[state,right of=0,minimum size=.5cm, label=below:{On}] (1) {};
	
	\path[->,thick, dashed] (0) edge [bend left] node {\it u\_on} (1);
	\path[->,thick, dashed] (1) edge [bend left] node {\it u\_off} (0);
	\end{tikzpicture}
\end{center}
		\caption{Sensor automaton} \label{fig:sensor}  
	\end{minipage}
\end{figure}  

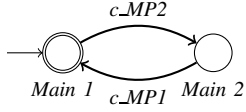
\begin{figure}[h]
	\begin{center}
	\begin{tikzpicture}[auto,->,node distance=2cm,align=center,font=\smaller\it, uncontrollable/.style={densely dashed}]
	\node[initial left, initial text = {}, state,accepting,minimum size=.5cm,label=below:{Main 1}] (0) {};
	\node[state,right of=0,minimum size=.5cm,label=below:{Main 2}] (1) {};
	
	\path[->,thick] (0) edge [bend left] node {\it c\_MP2} (1);
	\path[->,thick] (1) edge [bend left] node {\it c\_MP1} (0);
	\end{tikzpicture}
\end{center}
	\caption{Pump direction automaton}
	\label{fig:pdirection}	
\end{figure}

In Fig. \ref{fig:pump} and \ref{fig:sensor}, the automata of the models for the pumps and sensors are shown. Initial states are denoted by an arrow which does not originate in another state, double circles denote marked states, dashed and solid arrows denote uncontrollable and controllable events, respectively. The pump direction of the middle pump-cellar is modeled as shown in Fig. \ref{fig:pdirection}, in state $\mathit{Main 1}$ water is pumped towards main pump-cellar 1 and in state $\mathit{Main 2}$ water is pumped towards main pump-cellar 2. The mode model is shown in Fig. \ref{fig:mode}, event names are omitted here. 

\begin{figure}[h]  
	\centering 
	\begin{minipage}{0.49\linewidth}
		\begin{center}
	\begin{tikzpicture}[auto,->,node distance=3cm,align=center,font=\smaller\it, uncontrollable/.style={densely dashed}]
	\node[initial left, initial text = {}, state,accepting,minimum size=0.5cm, label=below:{Empty \ \quad}] (0) {};
	\node[state,draw,minimum size=0.5cm, label=above:{Store}] at (1.5,1) (1) {};
	\node[state,draw,minimum size=0.5cm, label=below:{Off}] at (1.5,-1) (2) {};
	
	\path[->,thick] (0) edge [bend left] 	node[sloped, above] {\it c\_store } (1);
	\path[->,thick] (1) edge [] 			node[sloped, below, yshift=.05cm] {\it c\_empty } (0);
	\path[->,thick] (0) edge [] 			node[sloped, above, yshift=-.05cm] {\it c\_off } (2);
	\path[->,thick] (2) edge [bend left] 	node[sloped, below, xshift=.1cm] {\it c\_empty } (0);
	\path[->,thick] (1) edge [bend left] 	node[sloped, above] {\it c\_off } (2);
	\path[->,thick] (2) edge [] 			node[sloped, above, yshift=-.05cm] {\it c\_store  } (1);
	\end{tikzpicture}
\end{center}
		\caption{Mode automaton} \label{fig:mode}	  
	\end{minipage}
	\begin{minipage}{0.49\linewidth}
		\begin{center}
	\begin{tikzpicture}[auto,->,node distance=3cm,align=center,font=\smaller\it, uncontrollable/.style={densely dashed}]
	\node[initial left, initial text = {}, state,accepting,minimum size=0.5cm, label=below:{Operational \quad }] (0) {};
	\node[state,minimum size=0.5cm, label=above:{Emergency}] at (1.5, 1) (1) {};
	\node[state,minimum size=0.5cm, label=below:{Recovery}] at (1.5, -1) (2) {};

	\path[->,thick] (0) edge [] 			node[sloped, above] {\it c\_em.} (1);
	\path[->,thick] (1) edge [bend left] 	node[sloped, above] {\it c\_rec.} (2);
	\path[->,thick] (2) edge [] 			node[sloped, above] {\it c\_em.} (1);
	\path[->,thick] (2) edge [] 			node[sloped, above] {\it c\_op.} (0);
	\end{tikzpicture}
\end{center}
		\caption{Traffic tube automaton} \label{fig:SV}  
	\end{minipage}
\end{figure}  

Each button is modeled as shown in Fig. \ref{fig:Button}, with a single event $u\_push$. Additionally for each set of buttons a monitor automaton is created. There are for example 3 buttons controlling the mode of a pump-cellar: $\mathit{ButtonEmpty}$, $\mathit{ButtonStore}$ and $\mathit{ButtonOff}$. The monitor automaton for these 3 buttons is shown in Fig. \ref{fig:Monitor}. In this automaton, the button events are observed. The requirements are stated such that $\mathit{Mode}$ events are allowed when the button monitor is in the relevant state. Event $\mathit{c\_done}$ is only allowed when $\mathit{Mode}$ has reached the relevant state. Other button sets are modeled in a similar fashion. 

\begin{figure}[h]  
	\centering 
	\begin{minipage}{0.39\linewidth}
		\begin{center}
	\begin{tikzpicture}[auto,->,node distance=3cm,align=center,font=\smaller\it, uncontrollable/.style={densely dashed}]
	\node[initial left, initial text = {}, state,accepting,minimum size=.5cm] (0) {};

	\path[->,thick, dashed] (0) edge [loop right] node {\it u\_push} (0);

	\end{tikzpicture}
\end{center}
		\caption{Button} \label{fig:Button}  
	\end{minipage}
	\begin{minipage}{0.59\linewidth}
		\begin{center}
	\begin{tikzpicture}[auto,->,node distance=2cm,align=center,font=\smaller\it, uncontrollable/.style={densely dashed}]
	\node[initial left, initial text = {}, state,accepting,minimum size=.5cm,label=below:{Idle}] (0) {};
	\node[state, above of= 0, minimum size = .5cm, label=above:{EmptyPushed}] (1) {};
	\node[state, minimum size = .5cm, label=below:{StorePushed}] at (2.3,0) (2) {};
	\node[state, below of= 0, minimum size = .5cm, label=below:{OffPushed}] (3) {};
	
	\path[->,thick, dashed] (0) edge [bend left] node {\it Empty.u\_push } (1);
	\path[->,thick] 		(1) edge [bend left] node {\it c\_done } (0);
	\path[->,thick, dashed] (0) edge [bend left] node {\it Store.u\_push } (2);
	\path[->,thick] 		(2) edge [bend left] node {\it c\_done  } (0);
	\path[->,thick, dashed] (0) edge [bend left] node {\it Off.u\_push } (3);
	\path[->,thick] 		(3) edge [bend left] node {\it c\_done  } (0);
	\end{tikzpicture}
\end{center}
		\caption{Button monitor automaton} \label{fig:Monitor}	 
	\end{minipage}
\end{figure}

\subsection{Requirement model}
A set of requirements is created to describe what behavior the controller should allow. For each pump-cellar, a number of requirements states when the pumps are allowed to turn on or off. When the $\mathit{Mode}$ of the pump-cellar is $\mathit{Empty}$, pump 1 is allowed to turn on when sensor level 2 is on and off when sensor level 1 is off. Pump 2 is allowed to turn on when sensor level 3 is on and off when sensor 1 is off. In $\mathit{Mode}$ $\mathit{Store}$, pumps 1 and 2 can both turn on when sensor level 5 is on and can both turn off when sensor level 4 is off. In $\mathit{Mode}$ $\mathit{Off}$ the pumps can only turn off. The middle pump-cellar initially pumps water towards main pump-cellar 1, the pump direction can be changed to main pump-cellar 2 when in main pump-cellar 1 sensor level 5 is turned on and can be changed back to main pump-cellar 1 if sensor level 4 is off.

All pump-cellars that are in automatic control mode are allowed to change their mode to $\mathit{Store}$ when a traffic tube is in state $\mathit{Emergency}$ or $\mathit{Recovery}$ and to $\mathit{Empty}$ when both traffic tubes are in state $\mathit{Operational}$. All pump-cellars that are in manual control mode, should change their mode according to the buttons. 

The requirements are stated as event conditions. Note that, these requirements can also be written as FA. An example of such a requirement is:\\

\textcolor{blue}{requirement} \textit{Pump1.c\_on} \textcolor{blue}{needs} not $\textit{Mode.Off}$;\\

This requirement states that event \textit{Pump1.c\_on} is only enabled if the $Mode$ automaton in not in the state $\mathit{Off}$. The requirements discussed in this section are all formulated in this manner. Note that, for 1 event, multiple requirements may exist. 

\section{STEP 1: Distributing the system}
\label{sec:Clustering}
The first step in acquiring a distributed supervisor is distributing the system into clusters. For each cluster a local supervisor can then be derived. Clusters can consist of for example events or requirements. In this paper, clusters of components are used, as this enables the use of multilevel synthesis based on this component clustering. The multilevel supervisor can then be localized to acquire local supervisors for each cluster of components. Moreover, clustering on component level gives the option of using a most refined product system (MRPS). Using the MRPS as a basis for clustering ensures that components that are closely connected, namely through shared events, are always in the same cluster.

After creating the MRPS for the system, a DSM is created for that MRPS. Different types of dependencies can be shown in a DSM. Dependencies can, for example, be defined based on the fysical location of components or electrical connections. Alternatively, a DSM can be created based on control relations, which is done in this paper, following \cite{Goorden2018}. This minimizes the amount of communication needed between local supervisors. To create such a DSM first Domain Mapping Matrices (DMM) $P1$ and $P2$ are created, which are rectangular matrices, see \cite{Wilschut2018}. These DMMs show relations between plant components and requirements in the system. In the CIF3 language, requirements are usually stated as follows:\\

\textcolor{blue}{requirement} \textit{event} \textcolor{blue}{needs} $\textit{condition}$;\\

Here, the \textit{event} is some event in the system and $\textit{condition}$ is a boolean expression which can, for example, refer to one or more states of one or more automata. $P1_{i,j} = 1$ when requirement $j$ refers to an event of component $i$ and $P1_{i,j} = 0$ otherwise. $P2_{i,j} = 1$ when the $\textit{condition}$ of requirement $j$ refers to an automaton of component $i$ and $P2_{i,j} = 0$ otherwise. The DSM $P$ can then be computed using $P=P1 \cdot P2^{T}$.

The acquired DSM can be clustered. Section \ref{sec:DSM} discusses a number of clustering methods. Each of these clustering methods can be used for clustering DSMs, however a number of characteristics of clustering algorithms are desirable. Firstly, control over the number of created clusters is needed, as for each cluster, a local supervisor will be synthesized. Secondly, the size of the created clusters needs to be controlled as this determines the number of components that need to be controlled by each local supervisor. Lastly, the algorithm must be able to cluster systems with a large number of components and dependencies. 

In K-means clustering, the number of clusters can be directly controlled, however the cluster size cannot be controlled. In spectral clustering, there is some control over the number of clusters, however creating a specific number of clusters can be inconvenient as the algorithm has to be used iteratively. The optimization methods are quite versatile, as control over the number of clusters and cluster size can be implemented. However, when analyzing large systems, control parameters of the algorithms need to be tuned very carefully. As the number of possible clusterings increases exponentially with the number of components, the computation times of the optimization algorithms increases greatly with the number of components. Markov clustering has been shown to be a scalable solution, where cluster size and the number clusters can be easily controlled through tuning the clustering parameters $\alpha$, $\beta$, $\mu$ and $\gamma$\cite{Wilschut2018}. In this paper, therefore, Markov clustering is used. 

%
%

\subsection{Case study: Pump-cellar system}
\begin{figure}
	\raggedright
	\includegraphics[width=\linewidth]{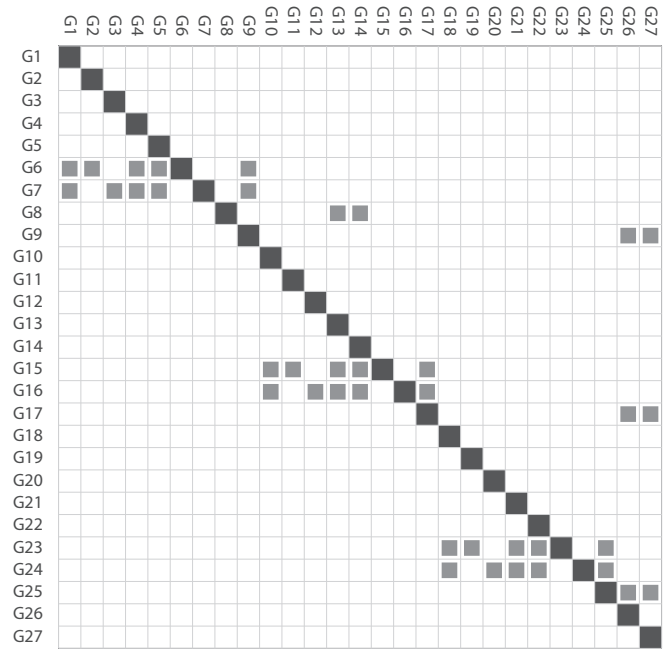}
	\caption{Unclustered pump-cellar DSM}
	\label{fig:unclustered}
\end{figure}

\begin{figure*}
	\raggedright
	\includegraphics[width=\linewidth]{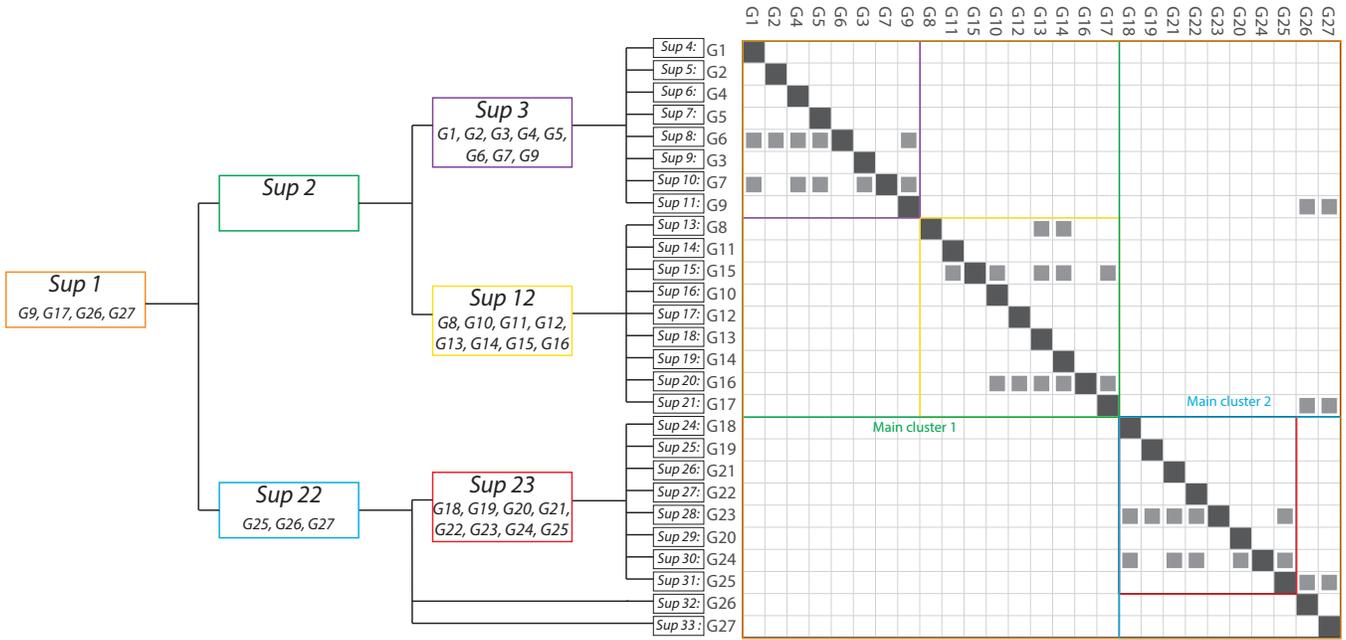}
	\caption{Multilevel supervisor tree and clustered pump-cellar DSM}
	\label{fig:clustered}
\end{figure*}

A DSM is created for the pump-cellar system introduced in Section \ref{sec:usecase}. First, the MRPS of the pump-cellar system is derived resulting in a system with 27 components and 37 requirements. This system is used to create DMMs $P1$ and $P2$. The resulting DSM is depicted in Fig. \ref{fig:unclustered}. The 27 components are listed at both axes as components $G_i$. Markov clustering is used to cluster the components in this DSM, with parameter tuning $\alpha = 2$, $\beta = 2.9$, $\mu = 2.9$, and $\gamma = 30$. For the application of localization, it is not desirable to isolate bus elements in a single cluster. Therefore, the chosen value for $\gamma$ is relatively high, as this eliminates the detection of bus elements. The resulting clustering is manually adjusted to the DSM depicted in Fig. \ref{fig:clustered}. 

As the distributed supervisor is to be implemented on two PLCs, two local supervisors, and therefore two main clusters are required. The first cluster (outlined by the green square in Fig. \ref{fig:clustered}) contains two subclusters, with the components of the middle pump-cellar and of the main pump-cellar 1, respectively. The second cluster (outlined by the blue square) contains the two traffic tubes and a cluster containing the components of main pump-cellar 2.

\section{STEP 2: DISTRIBUTING THE SUPERVISOR}
\label{sec:DistSynth}
After distributing the system, the second step is distributing the supervisor. First, multilevel synthesis is done according to \cite{Goorden2018}, using a plant model, a requirement model and the acquired distribution of the system. The tree structure, that is created in multilevel synthesis, is based on this distribution. Distributing the supervisor consists of three parts: the supervisor is localized, a delay-robustness check is done and, if needed, mutual exclusion algorithms are implemented.

\subsection{Supervisor Localization}
\label{sec:Localization}
The acquired multilevel supervisor consists of a set of supervisors similar to the example of Fig. \ref{fig:MLDES_example}. For each main cluster $k$, the set of relevant supervisors is taken as local supervisor $\textbf{LOC}_k$. This local supervisor can only disable events in the controllable alphabet $\Sigma_{c}^k$ of the components in its cluster. $\textbf{LOC}_k$ is adjusted such that all other events in $\Sigma \setminus \Sigma_{c}^k$ are always enabled, which is done as follows.

$\textbf{LOC}_k$ generally consists of a number of plant automata, requirements and one or more supervisor automata. First, requirements in the supervisor that disable events in $\Sigma \setminus \Sigma_{c}^k$ are removed from the supervisor. Next, any guards that disable events in $\Sigma \setminus \Sigma_{c}^k$ are removed. All plant automata of components outside of cluster $k$, that $\textbf{LOC}_k$ no longer refers to in its requirements or supervisor automata, are removed from $\textbf{LOC}_k$. If, after the adjustments, the supervisor automata or requirements in $\textbf{LOC}_k$ refer to events or states of automata that are not part of components in cluster $k$. Than that automaton is not removed from $\textbf{LOC}_k$, and it this paper, such a automaton is referred to as an observer. The events of observers in $\textbf{LOC}_k$ are not controlled by $\textbf{LOC}_k$.

For example, a requirement of $\textbf{LOC}_k$ refers to the state of an example automaton $A$, which is part of cluster $l$. As automaton $A$ is not in cluster $k$, $A$ is an observer in $\textbf{LOC}_k$. The occurrence of events of automaton $A$, is observed by $\textbf{LOC}_k$ in $\textbf{LOC}_l$, such that every time an event of $A$ happens in $\textbf{LOC}_l$, it is communicated to $\textbf{LOC}_k$. Note that, due to communication delays, events in observers are delayed. Events of $A$ are referred to as shared events.

\subsection{Delay-Robustness}
\label{sec:delayrobust}
\begin{figure}[h]
	\begin{center}
	\begin{tikzpicture}[auto,->,node distance=1cm,align=center,font=\smaller\it, uncontrollable/.style={densely dashed}]
	\node[initial left, initial text = {}, state,accepting,minimum size=0.5cm] (0) {\it 0};
	\node[state,below right = of 0,minimum size=.5cm] (1) {\it 1};
	\node[state,above right = of 0,minimum size=.5cm] (2) {\it 2};
	\node[state,right = of 1,accepting,minimum size=.5cm] (3) {\it 3};
	\node[state,right = of 2,accepting,minimum size=.5cm] (4) {\it 4};
	
	\path[->,thick] (0) edge [] node {\it a} (1);
	\path[->,thick] (2) edge [] node {\it a} (4);
	\path[->,thick] (0) edge [] node {\it b} (2);
	\path[->,thick] (1) edge [] node {\it b} (3);
	\path[->,thick] (3) edge [loop right] node {\it c} (3);
	\end{tikzpicture}
\end{center}
	\caption{Example local supervisor}
	\label{fig:dependent}
\end{figure}
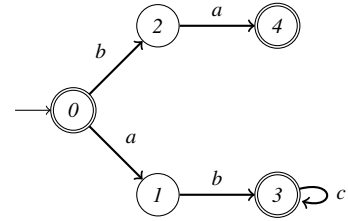

The local supervisors, derived by localization of the global supervisor, only impose globally correct behavior in the absence of communication delays. Communication delays might change the order in which events are observed in local supervisors. If, in the example local supervisor of Fig. \ref{fig:dependent}, event $b$ is delayed, the order in which events $a$ and $b$ occur might change. Due to this, event $c$ might be enabled where it should be disabled. Hence, communication delays might cause the localized supervisors to be globally unsafe, blocking, or uncontrollable.

As communication delays are unavoidable, distributed supervisors need to be adjusted to cope with these delays. \cite{Zhang2016} proposes a method to check if the distributed supervisor is delay robust with respect to events that are shared among local supervisors. A number of requirements is stated, for a distributed supervisor to be delay robust, i.e. a supervisor in which the behavior is not critically altered by communication delays. Within these requirements, equality of the languages of a supervisor with zero delay and a supervisor with finite delay, is guaranteed. In the CIF3 modeling language, guards and requirements often refer to the state of an automaton. Therefore, the requirements for delay robustness are defined, in this paper, such that local supervisors eventually reach the same state after delayed events. Definitions from \cite{Zhang2016}, in Equations \ref{eq:SUP}, \ref{eq:SUP'}, \ref{eq:2SUP} and \ref{eq:2SUP'} and Fig. \ref{fig:channel}, are used to define these requirements.

A global supervisor $\textbf{SUP}$ is defined by $n$ local supervisors: $\textbf{SUP} = ||_i \ \textbf{LOC}_i$, where $i = 1, 2, ..., n$. To check for delay robustness for a shared event in the supervisor, first, this shared event is renamed. Say event $r$ is shared between local supervisors $\textbf{LOC}_1$ and $\textbf{LOC}_2$. Event $r$ occurs in $\textbf{LOC}_1$ and is observed by $\textbf{LOC}_2$. $r$ is now renamed to $r'$ in $\textbf{LOC}_2$ creating $\textbf{LOC}'_2$. $r$, in $\textbf{LOC}_1$, is referred to as a channeled event and $r'$ is referred to as a delayed event. Next a channel is defined. Since $r$ is communicated from $\textbf{LOC}_1$ to  $\textbf{LOC}_2$, this channel is named $\textbf{CH}(1,r,2)$ and is defined as in Fig. \ref{fig:channel}. 

Using the channel and local supervisors both $\textbf{SUP}$ and  $\textbf{SUP}'$ can be defined as in Equations \ref{eq:SUP} and \ref{eq:SUP'}. $\textbf{SUP}$ can be seen as the distributed supervisor, whereas $\textbf{SUP}'$ is the distributed supervisor where communication delays are modeled.

\begin{figure}[h]
	\begin{center}
	\begin{tikzpicture}[auto,->,node distance=2cm,align=center,font=\smaller\it, uncontrollable/.style={densely dashed}]
	\node[initial left, initial text = {}, state,accepting,minimum size=0.5cm] (0) {};
	\node[state,right of=0,minimum size=0.5cm] (1) {};
	
	\path[->,thick] (0) edge [bend left] node {\it r} (1);
	\path[->,thick] (1) edge [bend left] node {\it r'} (0);
	\end{tikzpicture}
\end{center}
	\caption{$\textbf{CH}(1,r,2)$}
	\label{fig:channel}
\end{figure}

\begin{equation}
\label{eq:SUP}
\textbf{SUP} = || \ (\textbf{LOC}_1, \textbf{LOC}_2)
\end{equation}

\begin{equation}
\label{eq:SUP'}
\textbf{SUP}' = || \ (\textbf{LOC}_1, \textbf{CH}(1,r,2), \textbf{LOC}'_2)
\end{equation}

For $n$ local supervisors with multiple shared events, the definitions of Equations \ref{eq:SUP} and \ref{eq:SUP'} are adjusted. For each shared event, a channel, $\textbf{CH}(i,r,j)$ is defined, where $i$ is the index of the local supervisor where shared event $r$ is controlled and $j$ the index of the local supervisor which observes event $r$ from local supervisor $i$. If multiple channels exist for one channeled event, each channel has its own delayed event. The following definitions from \cite{Zhang2016} are used:

\begin{equation}
\label{eq:2SUP}
\textbf{SUP} = || \ (\textbf{LOC}_j | j \in N)
\end{equation}

\begin{equation}
\label{eq:2SUP'}
\textbf{SUP}' = || \ (\textbf{LOC}'_j, \textbf{CH}(i,r,j)| r \in \Sigma_{ch}(i,j), i \in I_j ,j \in N)
\end{equation}

Above, $N = \{1,2,...,n\}$, $I_j$ is the set of indexes of all supervisors from which $\textbf{LOC}_j$ observes events. $\Sigma_{ch}(i,j)$ is the set of all events which $\textbf{LOC}_j$ observes from $\textbf{LOC}_i$ and $\textbf{LOC}'_j$ is $\textbf{LOC}_j$ with all events in $\Sigma_{ch}(i,j)$ renamed. Note that if $I_j = \emptyset$ for $\textbf{LOC}_j$, i.e. if $\textbf{LOC}_j$ does not observe any events from other local supervisors, $\textbf{LOC}'_j = \textbf{LOC}_j$.

In this paper, sufficient conditions are defined to check if a distributed supervisor defined by FA is delay robust with respect to its shared events. First, the definitions of independence and mutual exclusivity, as defined by \cite{Reijnen2019a}, are given. 

Two events are called independent when they are enabled simultaneously and after the execution of either event, the other event is still enabled and either order reaches the same state. An example of this is shown in Fig. \ref{fig:independent}, where events $a$ and $b$ are independent. If two events are not enabled simultaneously, as shown in Fig. \ref{fig:ME} they are called mutually exclusive. 

\begin{figure}[h]  
	\centering 
	\begin{minipage}{0.49\linewidth}
	\begin{center}
	\begin{tikzpicture}[auto,->,node distance=1cm,align=center,font=\smaller\it, uncontrollable/.style={densely dashed}]
	\node[initial left, initial text = {}, state,accepting,minimum size=0.5cm] (0) {};
	\node[state,below right = of 0,minimum size=.5cm] (1) {};
	\node[state,above right = of 0,minimum size=.5cm] (2) {};
	\node[state,above right = of 1,accepting,minimum size=.5cm] (3) {};
	
	\path[->,thick] (0) edge [] node {\it a} (1);
	\path[->,thick] (2) edge [] node {\it a} (3);
	\path[->,thick] (0) edge [] node {\it b} (2);
	\path[->,thick] (1) edge [] node {\it b} (3);
	\end{tikzpicture}
\end{center}
	\caption{Independence} \label{fig:independent}	  
	\end{minipage}
	\begin{minipage}{0.49\linewidth}
	\begin{center}
	\begin{tikzpicture}[auto,->,node distance=1cm,align=center,font=\smaller\it, uncontrollable/.style={densely dashed}]
	\node[initial left, initial text = {}, state,accepting,minimum size=0.5cm] (0) {};
	\node[state,accepting,below right = of 0,minimum size=.5cm] (1) {};
	\node[state,accepting,above right = of 0,minimum size=.5cm] (2) {};
	
	\path[->,thick] (0) edge [] node {\it a when $x>0$} (1);

	\path[->,thick] (0) edge [] node[swap] {\it b  when $x<0$} (2);

	\end{tikzpicture}
\end{center}
	\caption{Mutual exclusiveness} \label{fig:ME} 
	\end{minipage}
\end{figure}

A number of assumptions is done in this paper, before defining the sufficient conditions for delay robustness. Firstly, it is assumed that communication occurs with a finite delay and can be modeled by a FIFO (First In First Out) queue. Secondly, it is assumed that an event $r$ is always followed by its delayed event $r'$ before a second occurrence of event $r$, as is modeled by the channels.

\textbf{Proposition 1:}\\
\textit{If every delayed event in $\textbf{SUP}'$ is either mutually exclusive with respect to all other events in $\textbf{SUP}'$ or independent with respect to all other simultaneously enabled events in $\textbf{SUP}'$, the distributed supervisor $\textbf{SUP}$ is delay robust with respect to all its delayed events.
}

A proof of Proposition 1 is provided in Appendix A. A distributed supervisor is said to be delay robust with respect to a single delayed event, if proposition 1 holds only for that event.

In \cite{Reijnen2019a}, a test is provided to check if a supervisor is confluent. Confluence is proven if, among other options, event combinations are independent or mutually exclusive. The existing checks for independence and mutual exclusiveness are used to check if delayed events are mutually exclusive or independent with all other events in $\textbf{SUP}'$, if this is the case the distributed supervisor $\textbf{SUP}$ is delay robust. 

It should be noted that communication channels are assumed to be FIFO queues, therefore if multiple events are communicated between two local supervisors, their order is preserved. Moreover, delayed versions of these events are mutually exclusive. Event combinations of two delayed events for which the original event is controlled by the same local supervisor, therefore need not be checked. 

\subsection{Mutual Exclusion}
If a supervisor is not delay robust with respect to a delayed event, this event is denoted a delay-critical event. A delay-critical event is not independent and not mutually exclusive with one or more events. The combinations of each of these events with the delay-critical event, are called delay-critical event combinations. When a distributed supervisor contains delay-critical event combinations, nonblockingness, safety, controllability cannot be guaranteed. Therefore the supervisor needs to be adjusted. There are two options, the first is to change the model of the system such that the distributed supervisor no longer contains delay-critical event combinations. This option is only possible in specific situations. Currently, no guidelines have been developed to identify such situations. The second option is in general less desirable, but it is always possible. Using mutual exclusion algorithms (mutex), it is possible to enforce the mutual exclusiveness of the delay-critical event combinations in the global supervisor $\textbf{SUP}$ as defined by equation \ref{eq:2SUP}. This then also implies that the event combinations are mutually exclusive in $\textbf{SUP}'$ and therefore the event combinations are no longer delay critical.

It should be noted that mutex algorithms cannot be used for uncontrollable events as it is not possible to disable uncontrollable events. Delay-critical event combinations containing uncontrollable events can currently not be handled. The pump-cellar case study, does not contain such event combinations.

An implementation of a mutex algorithm is referred to as a mutex lock. A delay-critical event combination of events $r'$ and $a$, can be made mutually exclusive by implementing a mutex lock for events $r$ and $a$. Event $r$ is controlled by a different local supervisor than event $a$. Each mutex lock therefore contains two processes, which are local supervisors in this implementation. Events $r$ and $a$ are always disabled by the corresponding local supervisor, unless it has entered the critical section. Therefore, $r$ and $a$ are mutually exclusive and consequently $r'$ and $a$ are mutually exclusive. Note that this is only true under the assumption that communication channels are FIFO queues. A home-based token passing algorithm, as discussed in Section \ref{sec:mutextheory}, is used, as this is a simple mutex algorithm, which is efficient for low numbers of processes.

For each mutex lock, a home process and an away process are defined. The mutex on the away process (referred to as process 2) is modeled using 3 automata and a number of requirements (Table \ref{tab:away}). The first automaton (Fig. \ref{fig:token}) models when the token is present, the second (Fig. \ref{fig:CS}) models the critical section and the third (Fig. \ref{fig:Requester}) models the token requesting procedure. The requirement of Fig. \ref{fig:Requirement} requires the token to be returned every time after using the critical section, this is done to prevent starvation of other processes. Requirements 5 and 7 contain a request condition and a return condition. Depending on the event ($critical\_event$) that is to be made mutually exclusive, these conditions state when the event is available (and therefore needs the token) and when the event has occurred (and therefore no longer needs the token), respectively.

\begin{table}[h]
	\caption{Requirements away process}
	\label{tab:away}
	\begin{tabular}{rllll}
		1 &\textcolor{blue}{requirement}  & $to\_home$        		& \textcolor{blue}{needs} & $CS.Idle$                                             \\
		2 &\textcolor{blue}{requirement}  & $to\_home$   		    & \textcolor{blue}{needs} & $Requester.Received$                                  \\
		3 &\textcolor{blue}{requirement}  & $to\_cs$                & \textcolor{blue}{needs} & $Token.Here$                                          \\
		4 &\textcolor{blue}{requirement}  & $to\_cs$                & \textcolor{blue}{needs} & $Requester.Received$                                  \\
		5 &\textcolor{blue}{requirement}  & $to\_idle$              & \textcolor{blue}{needs} & $ReturnCondition$                                     \\
		6 &\textcolor{blue}{requirement}  & $request$               & \textcolor{blue}{needs} & \textcolor{cyan}{not} $Token.Here$ \\
		7 &\textcolor{blue}{requirement}  & $request$               & \textcolor{blue}{needs} & $RequestCondition$                                    \\
		8 &\textcolor{blue}{requirement}  & $received$              & \textcolor{blue}{needs} & $Token.Here$                                          \\
		9 &\textcolor{blue}{requirement}  & $return$                & \textcolor{blue}{needs} & \textcolor{cyan}{not} $Token.Here$ \\
		10 &\textcolor{blue}{requirement} & $critical\_event$		& \textcolor{blue}{needs} & $CS.Active$                                          
	\end{tabular}

\end{table}

\begin{figure}[h]  
	\centering 
	\begin{minipage}{0.49\linewidth}
	\begin{center}
	\begin{tikzpicture}[auto,->,node distance=2cm,align=center,font=\smaller\it, uncontrollable/.style={densely dashed}]
	\node[initial left, initial text = {}, state,accepting,minimum size=0.5cm, label=below:{NotHere}] (0) {};
	\node[state,right of=0,minimum size=0.5cm, label=below:{Here}] (1) {};
	
	\path[->,thick] (0) edge [bend left] node {\it to\_here} (1);
	\path[->,thick] (1) edge [bend left] node {\it to\_home} (0);
	\end{tikzpicture}
\end{center}
	\caption{Token} \label{fig:token} 
	\end{minipage}
	\begin{minipage}{0.49\linewidth}
 	\begin{center}
	\begin{tikzpicture}[auto,->,node distance=2cm,align=center,font=\smaller\it, uncontrollable/.style={densely dashed}]
	\node[initial left, initial text = {}, state,accepting,minimum size=0.5cm, label=below:{Idle}] (0) {};
	\node[state,right of=0,minimum size=0.5cm, label=below:{Active}] (1) {};
	
	\path[->,thick] (0) edge [bend left] node {\it to\_cs} (1);
	\path[->,thick] (1) edge [bend left] node {\it to\_idle} (0);
	\end{tikzpicture}
\end{center}
	\caption{Critical Section (CS)} \label{fig:CS} 
	\end{minipage}
\end{figure}  

%
%

\begin{figure}[h]  
	\centering 
	\begin{minipage}{0.49\linewidth}
	\begin{center}
	\begin{tikzpicture}[auto,->,node distance=2cm,align=center,font=\smaller\it, uncontrollable/.style={densely dashed}]
	\node[initial left, initial text = {}, state,accepting,minimum size=0.5cm, label=below:{Idle}] (0) {};
	\node[state,minimum size=0.5cm, label=above:{Requested}] at (2, 1) (1) {};
	\node[state,minimum size=0.5cm, label=below:{Received}] at (2, -1) (2) {};
		
	\path[->,thick] (0) edge [] node[sloped, above] {\it request} (1);
	\path[->,thick] (1) edge [] node[sloped, above] {\it received} (2);
	\path[->,thick] (2) edge [] node[sloped, above] {\it return} (0);
	\end{tikzpicture}
\end{center}
	\caption{Requester}
	\label{fig:Requester}
	\end{minipage}
	\begin{minipage}{0.49\linewidth}
	\begin{center}
	\begin{tikzpicture}[auto,->,node distance=2cm,align=center,font=\smaller\it, uncontrollable/.style={densely dashed}]
	\node[initial left, initial text = {}, state,accepting,minimum size=0.5cm,label = below:{A} ] (0) {};
	\node[state,minimum size=0.5cm,label = above:{B}] at (0, 2) (1) {};	
	\node[state,right of=1,minimum size=0.5cm,label = above:{C}] (2) {};
	\node[state,right of=0,minimum size=0.5cm,label = below:{D}] (3) {};
			
	\path[->,thick] (0) edge [] node[sloped, above] {\it to\_here} (1);
	\path[->,thick] (1) edge [] node {\it to\_cs} (2);
	\path[->,thick] (2) edge [] node[sloped, above] {\it to\_idle} (3);
	\path[->,thick] (3) edge [] node {\it to\_home} (0);
	\end{tikzpicture}
\end{center}
	\caption{Req. automaton}
	\label{fig:Requirement}
	\end{minipage}
\end{figure}  
 
%

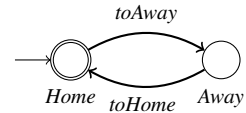
\begin{figure}[h]
	\begin{center}
	\begin{tikzpicture}[auto,->,node distance=2cm,align=center,font=\smaller\it, uncontrollable/.style={densely dashed}]
	\node[initial left, initial text = {}, state,accepting,minimum size=0.5cm, label=below:{Home}] (0) {};

	\node[state, right of=0,minimum size=0.5cm, label=below:{Away}] (1){};

	\path[->,thick] (0) edge [bend left] node {\it toAway} (1);
	\path[->,thick] (1) edge [bend left] node {\it toHome } (0);	
	
	\end{tikzpicture}
\end{center}
	\caption{Token Tracker}
	\label{fig:tracker}
\end{figure}

The home process (process 1) consists of two automata and 5 requirements. The first automaton is the token tracker, as shown in Fig. \ref{fig:tracker}. This automaton tracks if the token is at the home or the away process. The second is the critical section automaton, which is the same as the critical section automaton of the away process shown in Fig. \ref{fig:CS}. An input boolean $R2$ is true if the away process is requesting the token, i.e. $R2$ is set to true if the away process \textit{Requester} automaton is in the state \textit{Requested}. The requirements of the home process are listed in Table \ref{tab:home}. The requirements state that the home process can enter the critical section when the token is at the home process and the $RequestCondition$ is true. Note that, different from the away process, the token does not need to be requested first. This implementation of a home-based token passing algorithm is not starvation free. The home process can keep entering the critical section while the away process is requesting the token. A queue automaton can be added to acquire starvation freedom, however, this increases the statespace of local supervisors. In this paper, such a queue automaton is not used, with the aim of reducing the statespace of local supervisors.

\begin{table}[h]
	\caption{Requirements home process}
	\label{tab:home}
	\begin{tabular}{rllll}
		1 &\textcolor{blue}{requirement} & $token\_to\_Away$        & \textcolor{blue}{needs} & $CS.Idle$            	\\
		2 &\textcolor{blue}{requirement} & $token\_to\_Away$        & \textcolor{blue}{needs} & $R2$  					\\
		3 &\textcolor{blue}{requirement} & $to\_cs$                 & \textcolor{blue}{needs} & $Tracker.Home$          \\
		4 &\textcolor{blue}{requirement} & $to\_cs$                 & \textcolor{blue}{needs} & $RequestCondition$      \\
		5 &\textcolor{blue}{requirement} & $to\_idle$               & \textcolor{blue}{needs} & $ReturnCondition$       \\
	\end{tabular}
\end{table}

As a mutex lock is implemented for each delay-critical event combination, it is possible that an event is the critical event in multiple mutex locks. \cite{Silberschatz2009} states that deadlock can occur if multiple mutex locks are used with an overlapping critical section. When two processes contain two mutex locks, mutex lock $A$ and mutex lock $B$, deadlock occurs when process 1 contains the token of lock $A$ and requires the token of lock $B$ and process two contains the token of lock $B$ and requires the token of lock $A$. To prevent deadlock, the mutex locks are ordered such that a process can only acquire tokens from mutex locks in that order, as discussed by \cite{Silberschatz2009}. The order in which tokens are acquired is controlled by adjusting the condition $RequestCondition$, such that a lock can only be acquired if relevant locks earlier in the order are already acquired.

Using mutex locks increases the statespace of local supervisors. In order to lower the statespace, it is possible to use a slightly different implementation of the mutex locks. The choice can be made to use the mutex locks, as described in this section, for two sets of events instead of using it for two events. Instead of using one $critical\_event$ in each process of the mutex, a critical event set for each process is used. All events of the home process critical event set are then mutually exclusive with all events of the away process critical event set.

\subsection{Case study: Pump-cellar system}
In Section \ref{sec:Clustering}, a clustering for the pump-cellar system is determined. Based on this clustering multilevel synthesis is done. The multilevel tree is depicted in Fig. \ref{fig:clustered}. The tree contains supervisors for each cluster, as well as a supervisor for each individual component. For every supervisor in the tree, the relevant components are shown, the requirements are omitted, to keep te figure simple. The resulting multilevel supervisor consists of a set of 33 supervisors. Note that supervisor 2 does not contain any components or requirements. Due to the chosen clustering, there are no requirements that refer to components of both the subclusters of main cluster 1. Therefore, all requirements that only refer to components of main cluster 1, are in $Sup3$ or $Sup12$. The advantage of using multilevel synthesis as a basis for localization, is that only supervisor $Sup 1$ refers to components of both the main clusters, all other supervisors only refer to components of one of the main clusters. 

The first step in distributing this multilevel supervisor is localization. As the distributed supervisor is implemented on two PLCs, two local supervisors, $\textbf{LOC}_1$ and $\textbf{LOC}_2$, are created. $\textbf{LOC}_1$ is created using the set of supervisors 1-21, where $Sup 1$ is the only supervisor that needs to be adjusted. $\textbf{LOC}_2$ is created using supervisors 1 and 22-33, where again only $Sup 1$ needs to be adjusted. $Sup 1$ adjusted for $\textbf{LOC}_1$ is referred to as $Sup 1_a$ and $Sup1$ adjusted for $\textbf{LOC}_2$ is referred to as $Sup 1_b$. $\textbf{LOC}_1$ is the set of supervisors 2-21 and the adjusted supervisor $Sup 1_a$. $\textbf{LOC}_2$ is the set of supervisors 22-33 and the adjusted supervisor $Sup 1_b$.

When $Sup 1$ is adjusted to $Sup 1_a$, the requirements and guards that disable events in components $G26$ and $G27$ are removed. The adjusted supervisor does not disable any events from component $G26$ and $G27$. The remaining requirements still refer to components $G26$ and $G27$, therefore $G26$ and $G27$ are not removed from $Sup 1_a$. For clarity, they are renamed, such that it is clear the automata of these components are observers.

To create $Sup 1_b$, all requirements and guards in $Sup 1$ that disable events in components $G9$ and $G17$ are removed. $Sup 1_b$ does not refer to components $G9$ and $G17$, therefore these components are removed from $Sup 1_b$.

Now that the supervisors are localized, the next step is to check for delay robustness. As $\textbf{LOC}_1$ contains observer components $G26$ and $G27$, the events of these components are delayed events in $\textbf{LOC}_1$. The delay robustness check discussed in Section \ref{sec:delayrobust} is used to check if the distributed supervisor is delay robust with respect to these delayed events. The check indicates a set of 36 event combinations which are neither independent nor mutually exclusive. 

The final step in creating a distributed supervisor is setting up mutexes for of these 36 event combinations. A number of event sets is chosen such that the 36 event combinations can be made mutually exclusive, using 4 mutex locks. An example of a delay-critical event combination is event $c\_operational'$ in the first traffic tube component ($G26$) and event $c\_store$ in the $Mode$ automaton of the middle pump-cellar ($G9$). Event $c\_operational$ is controlled by $\textbf{LOC}_2$ and event $c\_store$ by $\textbf{LOC}_1$. Event $c\_operational$ is in the critical event set of the home process of mutex lock 1 and $c\_store$ is in the critical event set of the away process of mutex lock 1. Therefore, these events are now mutually exclusive. For each of the 36 event combinations a mutex lock is used in a similar fashion. The resulting distributed supervisor is delay robust, nonblocking, safe and controllable.

\section{Hardware in the Loop testing}
\label{sec:HIL}
A Hardware in the Loop (HIL) setup was used to test the distributed supervisor. Before implementing the supervisor on a PLC, some adjustments need to be made. \cite{Reijnen2020a} denotes three properties that a supervisor needs to adhere to when a controller is derived from that supervisor. These properties are confluence, finite response, and nonblockingness under control. In CIF3 checks are developed to verify if a supervisor has each of these properties.

Confluence is defined as follows by \cite{Malik2003}. Whenever a controller can choose between two controllable events, each of these events can be extended by a string of controllable events such that both paths end up in the same state. A controller has finite response when it does not contain an infinite sequence of controllable events. Finally, \cite{Malik2003} defines nonblockingness under control. From every reachable state in a controller, a sequence of events, which prioritizes controllable events over uncontrollable event, is enabled and reaches a marked state. Moreover, in this marked state no controllable events are enabled. If this is the case, the controller is nonblocking under control. \cite{Reijnen2020a} discusses these properties and the developed checks in more detail.

\begin{figure}[h]
	
\begin{center}
	\begin{tikzpicture}[auto,->,node distance=1.5cm,align=center,font=\smaller\it, uncontrollable/.style={densely dashed}]
	\node[initial above, initial text = {}, state,accepting,minimum size=.5cm, label=below:{Off}] (0) {};
	\node[state,right of=0,minimum size=.5cm, label=below:{L 1}] (1) {};
	\node[state,right of=1,minimum size=.5cm, label=below:{L 2}] (2) {};
	\node[state,right of=2,minimum size=.5cm, label=below:{L 3}] (3) {};
	\node[state,right of=3,minimum size=.5cm, label=below:{L 4}] (4) {};
	\node[state,right of=4,minimum size=.5cm, label=below:{L 5}] (5) {};
	
	\path[->,thick, dashed] (0) edge [bend left] node {\it s1\_on} (1);
	\path[->,thick, dashed] (1) edge [bend left] node {\it s1\_off} (0);
	\path[->,thick, dashed] (1) edge [bend left] node {\it s2\_on} (2);
	\path[->,thick, dashed] (2) edge [bend left] node {\it s2\_off} (1);
	\path[->,thick, dashed] (2) edge [bend left] node {\it s3\_on} (3);
	\path[->,thick, dashed] (3) edge [bend left] node {\it s3\_off} (2);	
	\path[->,thick, dashed] (3) edge [bend left] node {\it s4\_on} (4);
	\path[->,thick, dashed] (4) edge [bend left] node {\it s4\_off} (3);
	\path[->,thick, dashed] (4) edge [bend left] node {\it s5\_on} (5);
	\path[->,thick, dashed] (5) edge [bend left] node {\it s5\_off} (4);
	\end{tikzpicture}
\end{center}

	\caption{Adjusted sensors automaton}
	\label{fig:sensors2}	
\end{figure}
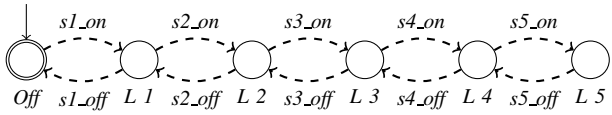

The controller checks are performed for the pump-cellar system. Some changes are required to achieve the three controller properties. Initially the sensors of each pump-cellar are modeled as individual sensors with no physical relations. This model is changed to one sensor automaton for all sensors in a pump-cellar. The automaton is given in Fig. \ref{fig:sensors2}, the event and state names in this figure are abbreviated. In the initial model, the assumption that, for example, sensor level 1 would never be off if sensor level 2 is on. In the adjusted model, this is incorporated. A new distributed supervisor is acquired for the adjusted model, in the same way as is described earlier. The resulting supervisor has the three controller properties.

The statespace of the supervisors is given in Table \ref{tab:statespace}. A significant reduction of the statespace is achieved by distributing the supervisor.

\begin{table}[h]
	\caption{Statespace of supervisors}
	\label{tab:statespace}
	\begin{tabular}{lll}
		& Statespace                 & \% of global \\
		Global multilevel supervisor & $54.4*10^9$ & 100          \\
		Local supervisor 1           & $98.0*10^8$ & 18           \\
		Local supervisor 2           & $77.6*10^6$ & 0.14        
	\end{tabular}
\end{table}

The distributed supervisor is implemented on two PLCs, which communicate using industrial ethernet. Before implementing the distributed supervisor, a hardware mapping is created, which maps events to input and output variables. The pump-cellars are simulated using Ignition Inductive Automation software, which is also used to create a graphical user interface consisting of all the buttons and a visualization of sensor values. A set of scenarios is defined to validate the behavior of the controlled system.  

\section{CONCLUDING REMARKS}
\label{sec:Conclusions}
In this paper, a method for developing distributed supervisory controllers is proposed. The resulting distributed supervisor is delay robust, nonblocking, controllable and safe. The method is demonstrated by developing a supervisor for a pump-cellar system. The developed supervisor is tested using simulation and a HIL-setup, validating the behavior of the controlled system. 

The statespace of the local supervisors is significantly lower than the statespace of the global supervisor. Which indicates that the distributed supervisor will require less memory and can achieve lower cycle times, than the global multilevel supervisor. 

The proposed method uses the top-down approach of localization, for which synthesis of a global supervisor is required. For large scale systems this might prove to be impossible. Therefore, more research is needed to investigate bottom-up methods, as for example has been proposed by \cite{Su2010}.

In the method of this paper, the statespace of the supervisor is greatly increased by the implementation of mutex locks. One of the reasons for using distributed supervisory controllers, is being able to control large scale systems, which cannot be controlled by other supervisory controllers. The increase of the statespace by mutex locks counteracts the ability to control large-scale systems, using the method of this paper.. Using a lower number of mutex locks is a solution to the statespace problem, but it is unclear what the effects are of using a single mutex lock for a large set of events. Therefore, more research is needed to more efficiently deal with delay-critical event combinations.



\bibliographystyle{unsrt}
\bibliography{Distributed_supervisor_synthesis.bbl}

\section*{APPENDIX A}
\label{sec:appA}
\subsection*{Proof of proposition 1:}
Recall proposition 1:\\
\textit{If every delayed event in $\textbf{SUP}'$ is either mutually exclusive with respect to all other events in $\textbf{SUP}'$ or independent with respect to all other simultaneously enabled events in $\textbf{SUP}'$, the distributed supervisor $\textbf{SUP}$ is delay robust with respect to all its delayed events.
}\\

Recall that, in a channel, $r$ is referred to as the channeled event and $r'$ is referred to as the delayed event. Furthermore, $\textbf{SUP} = (X, \Sigma, \xi , x_0,X_m)$ and $\textbf{SUP}' = (X', \Sigma', \xi' , x'_0,X'_m)$.\\

Firstly, note that by definition of $\textbf{SUP}'$, in a state reached by channeled event $r$,  delayed event $r'$ is always enabled.\\

Secondly, if $r'$ is mutually exclusive or independent with all other events in $\Sigma'^*$, it is said that $r'$ is independent with its simultaneously enabled events.\\

Thirdly, note that an event $r$ is only enabled after a string $s \in L(\textbf{SUP})'$, if for every event $r$ in $s$, $s$ contains an event $r'$, due to the definition of the channels. Subsequently, $r$ is only enabled after $s$ if $s$ contains the same number of events $r$ and $r'$.\\ 

Fourthly, note that $(\forall s \in L(\textbf{SUP}'))$ in which every channeled event is directly followed by its delayed event, $P(s) \in L(\textbf{SUP})$. This is true, as essentially no delay has occurred on any event. Such a string $s$ is referred to as a delay-free string, denoted by subscript $s_{df}$. Similarly, $(\forall s \in L_m(\textbf{SUP}'))$ in which every channeled event is directly followed by its delayed event, $P(s) \in L_m(\textbf{SUP})$.\\

Fifthly, note that for any string $s$, where any channeled event is at some point followed by its delayed event, there exists a string $s_{df}$ such that $P(s) = P(s_df)$, moreover, $\xi'(x_0,s) = \xi'(x_0,s_{df})$. This is true as any delayed event is independent with its simultaneously enabled events. For example, consider a string \\ 
$(s_1rs_2r's_3 \in L(\textbf{SUP}') \ | \ s_1 \in \Sigma^*, s_2 \in \Sigma^*, s_3 \in \Sigma^*)$. Note that $P(s_1rs_2r's_3) = s_1rs_2s_3 = P(s_1rr's_2s_3)$, moreover, as $r'$ is independent with its simultaneously enabled events, $\xi'(x_0,s_1rs_2r's_3) = \xi'(x_0,s_1rr's_2s_3)$. Here, $s_1rr's_2s_3$ is a delay free string as any channeled event, i.e. $r$, is directly followed by its delayed event, in this case $r'$.\\

$\textbf{SUP}$ is said to be delay robust with respect to its channeled events if the following 5 conditions, defined by \cite{Zhang2016}, hold:

\begin{equation}
\label{eq:cond1}
P L\left(\mathbf{S U P}^{\prime}\right) \subseteq L(\mathbf{S U P})
\end{equation}

\begin{equation}
\label{eq:cond2}
P L_{m}\left(\mathbf{S U P}^{\prime}\right) \subseteq L_{m}(\mathbf{S U P})
\end{equation}

\begin{equation}
\label{eq:cond3}
P L\left(\mathbf{S U P}^{\prime}\right) \supseteq L(\mathbf{S U P})
\end{equation}

\begin{equation}
\label{eq:cond4}
P L_{m}\left(\mathbf{S U P}^{\prime}\right) \supseteq L_{m}(\mathbf{S U P})
\end{equation}

\begin{equation}
\label{eq:cond5}
\begin{aligned}
\left(\forall a \in \Sigma^{\prime *}\right)\left(\forall c \in \Sigma^{*}\right) \\
 \quad a \in L\left(\mathrm{\textbf{SUP}}^{\prime}\right) \ \ \& \ \ P(a).c \in L_{m}(\mathrm{\textbf{SUP}}) \\
 \Rightarrow\left(\exists b \in \Sigma^{\prime *}\right) \ \ P(b)=c \ \ \& \ \ a.b  \in L_{m}\left(\mathrm{\textbf{SUP}}^{\prime}\right)
\end{aligned}
\end{equation} 

Each of these conditions is proven in this appendix, therefore condition 1 holds.

\subsection{Condition 1 (equation \ref{eq:cond1})}
It is proven by induction that:

$( \forall s \in L(\textbf{SUP}')) Ps \in L(\textbf{SUP})$.\\

Base step: $\epsilon \in L(\textbf{SUP}')$ and $\epsilon \in L(\textbf{SUP})$, trivially. 

Inductive step: suppose $t \in L(\textbf{SUP})'$, $P(t) \in L(\textbf{SUP})$ and  
$t.a \in L(\textbf{SUP}')$, we must prove that  $P(t.a) \in L(\textbf{SUP})$.\\

Two possible cases are identified:
\begin{enumerate}
	\item $t$ contains a delayed event for every occurrence of a channeled event. (i.e. no delayed event is enabled in $\xi'(x_0,t)$.)
	\item $t$ does not contain a delayed event for every occurrence of a channeled event. (i.e. at least one delayed event is enabled in $\xi'(x_0,t)$.)\\
\end{enumerate}

In case 1), $(\exists t_{df} \in L(\textbf{SUP}') \ | \ \xi'(x_0, t_{df}) = \xi'(x_0, t))$, where $t_{df}$ is a delay free string. In this case, $a$ cannot be a delayed event, as every delayed event is disabled by its channel, i.e. $t.a \notin L(\textbf{SUP}')$. If $a$ is a channeled event, $t_{df}.a.a' \in L(\textbf{SUP}')$ is again a delay free string as every channeled event is directly followed by its delayed event. Now, $P(t.a) = P(t_{df}.a) = P(t_{df}.a.a') \in L(\textbf{SUP})$. If $a$ is not a channeled event, $t_{df}.a$ is a delay free string. Therefore, $P(t.a) = P(t_{df}.a) \in L(\textbf{SUP})$.\\

In case 2), $t.a$ can be extended with a string of delayed events $d$, such that the string $t.a.d \in L(\textbf{SUP}')$ contains a delayed event for every occurrence of a channeled event. Therefore, $P(t.a.d) = P(t.a)$ and \\
$(\exists t_{df} \in L(\textbf{SUP}') \ | \ \xi'(x_0, t_{df}) = \xi'(x_0, t.a.d))$, where $t_{df}$ is a delay free string. Now, \\
$P(t.a) = P(t.a.d) = P(t_{df}) \in L(\textbf{SUP})$.\\

In both the cases, $(\forall a \in \Sigma' \ | \ t.a \in L(\textbf{SUP}'))$ it is proven that $P(t.a) \in L(\textbf{SUP})$, therefore condition 1 of equation \ref{eq:cond1} holds.

\subsection*{Condition 2 (equation \ref{eq:cond2})}
Condition 2 holds if: $( \forall s \in L_m(\textbf{SUP}')) P(s) \in L_m(\textbf{SUP})$.\\

Any string $s$ contains a delayed event for every occurrence of a channeled event, by definition, otherwise $s$ does not reach a marked state due to the definition of the channels. Therefore,  $(\forall s \in L_m(\textbf{SUP}')) \ (\exists s_{df} \ | \ \xi'(x_0, s_{df}) = \xi'(x_0, s))$, where $s_{df}$ is a delay free string. $P(s) = P(s_{df}) \in L_m(\textbf{SUP})$ by definition of delay-free strings.

\subsection*{Condition 3 (equation \ref{eq:cond3})}
Condition 3 holds if: $\left( \forall s \in L(\textbf{SUP})  \right) \ \ s \in PL(\textbf{SUP}')$\\

For any string $s$, a delay free string $s_{df} \in L(\textbf{SUP}')$ can be created by replacing any channeled event in $s$ by the channeled event directly followed by its delayed event. For example, if $s$ contains a channeled event $r$, create $s_{df}$ by replacing every occurrence of $r$ by $rr'$. By definition, $\textbf{SUP}'$ must allow such a delay free string.\\ 

As $s_{df} \in L(\textbf{SUP}')$, it must hold that \\
$s = P(s_{df}) \in PL(\textbf{SUP}')$.

\subsection*{Condition 4 (equation \ref{eq:cond4})}
Condition 4 holds if:

$\left( \forall s \in L_m(\textbf{SUP})  \right) \ \ s \in PL_m(\textbf{SUP}')$.\\

For any string $s$, a delay free string $s_{df} \in L_m(\textbf{SUP}')$ can be created by replacing any channeled event in $s$ by the channeled event directly followed by its delayed event, as is done for Condition 3. By definition, $\textbf{SUP}'$ must allow such a delay free string. \\

As $s_{df} \in L_m(\textbf{SUP}')$, it must hold that\\
 $s = P(s_{df}) \in PL_m(\textbf{SUP}')$.

\subsection*{Condition 5 (equation \ref{eq:cond5})}

Two possible cases are defined:
\begin{itemize}
	\item $a$ contains a delayed event for every occurrence of a channeled event. (i.e. no delayed event is enabled in $\xi'(x_0,t)$.)
	\item $a$ does not contain a delayed event for every occurrence of a channeled event. (i.e. at least one delayed event is enabled after in $\xi'(x_0,t)$.)\\
\end{itemize}

In case 1), due to the independence of delayed events with their simultaneously enabled events, $(\exists a_{df} \in L(\textbf{SUP}')\ | \ \xi'(x_0, a_{df}) = \xi'(x_0, a))$, where $a_{df}$ is a delay free string. Next, for any string $c$ a delay free string $c_{df}$ can be created as is done for Condition 3. Let $b=c_{df}$, then $P(b) = c$. Moreover, as $a_{df}.b$ is a delay free string and $P(a_{df}.b) = P(a).c \in L_m(\textbf{SUP})$, it must hold that $a_{df}.b \in L_m(\textbf{SUP}') \Rightarrow a.b \in L_m(\textbf{SUP}')$.\\

In case 2), $a$ can be extended with a string of delayed events $d$, such that the string $a.d \in L(\textbf{SUP}')$ contains a delayed event for every occurrence of a channeled event. For $a.d$ it must hold that \\
$(\exists a_{df} \in L(\textbf{SUP}')\ | \ \xi'(x_0, a_{df}) = \xi'(x_0, a.d))$, due to the independence of delayed events with their simultaneously enabled events. Next, for any string $c$ a delay free string $c_{df}$ can be created as is done for Condition 3. Let\\ 
$b = d.c_{df}$, then $P(b) = c$. Due to the independence of delayed events with their simultaneously enabled events, $(\exists f_{df} \in L(\textbf{SUP}')\ | \ \xi'(x_0, f_{df}) = \xi'(x_0, a_{df}.c_{df}) = \xi'(x_0, a.d.c_{df})$. As $f_{df}$ is a delay free string and \\
$P(f_{df}) = P(a).c \in L_m(\textbf{SUP})$, it must hold that\\ $f_{df} \in L_m(\textbf{SUP}') \Rightarrow a.b \in L_m(\textbf{SUP}')$.

\end{document}